\documentclass[graybox]{svmult}

\usepackage{mathptmx}       % selects Times Roman as basic font
\usepackage{helvet}         % selects Helvetica as sans-serif font
\usepackage{courier}        % selects Courier as typewriter font
\usepackage{makeidx}         % allows index generation
\usepackage{graphicx}        % standard LaTeX graphics tool
\usepackage{multicol}        % used for the two-column index
\usepackage[bottom]{footmisc}% places footnotes at page bottom
\usepackage[applemac]{inputenc}
\usepackage[T1]{fontenc}
\usepackage{xspace}
\usepackage[raggedright]{sidecap}
\usepackage{multicol,multirow,dcolumn,colortbl,lscape}
\usepackage{graphicx}
\usepackage{pdfpages}
\includepdfset{pages=-,link,pagecommand={\thispagestyle{fancy}}}
\usepackage[svgnames]{xcolor}
\usepackage{amsmath,amssymb}
\usepackage{booktabs}

\usepackage{natbib}

\makeindex             % used for the subject index
\usepackage{enumitem}
\setlist[description,1]{font=\bfseries, itemsep=8pt}
\newcommand{\ie}{i.\thinspace{}e.\xspace}

\begin{document}

\title*{Models of rotating massive stars: impacts of various prescriptions}
\author{Georges Meynet, Sylvia Ekstrom, Andr\'e Maeder, Patrick Eggenberger, Hideyuki Saio, Vincent Chomienne, Lionel Haemmerl\'e}
\authorrunning{Impacts of various prescriptions} 
%for an abbreviated version of
\newcommand{\gobs}{%
  Geneva Observatory, Geneva University, CH-1290 Versoix, Switzerland\xspace
  \renewcommand{\gobs}{Geneva Observatory, \textit{ibid.}\xspace}
}
\institute{Georges Meynet \at \gobs\\ \email{georges.meynet@unige.ch}
\and
Sylvia Ekstrom \at \gobs \email{Sylvia.Ekstrom@unige.ch}
\and
Andr\'e Maeder \at \gobs \email{Andr\'e.Maeder@unige.ch}
\and
Patrick Eggenberger \at \gobs \email{Patrick.Eggenberger@unige.ch}
\and 
Hideyuki Saio \at Astronomical Institute, Graduate School of Science, Tohoku University, Sendai 980-8578, Japan \email{saio@astr.tohoku.ac.jp}
\and
Vincent Chomienne \at  \gobs \email{Vincent.Chomienne@unige.ch}
\and
Lionel Haemmerl\'e \at \gobs \email{Lionel.Haemmerle@unige.ch}
}
\maketitle

\abstract{The rotation of stars has many interesting 
and important consequences for the photometric and chemical evolution 
of galaxies. Many of the predictions of models of stellar rotation are now compared with observations of surface abundances and velocities, 
with interferometric studies of fast rotating stars, with internal rotation profiles 
as they can be deduced by asteroseismology, to cite 
just a few observational constraints. 
In this paper, we investigate how the outputs of models depend on the prescriptions used
for the diffusion coefficients included in the 
shellular rotating models. After recalling the various 
prescriptions found in the literature, we discuss their impacts on the evolutionary 
tracks and lifetimes of the Main-Sequence (MS) phase, 
the changes of the surface composition and velocities during the MS phase,
the distribution of the core helium lifetime in the blue and the red part of the HR diagram,
the extensions of the blue loops,
the evolution of the angular momentum of the core,
and the synthesis of primary nitrogen in fast-rotating metal-poor
massive stars. 
While some of these outputs depend only slightly
on the prescriptions used (for instance, the evolution of the surface 
velocities), most of them show
a significant dependence. 
The models which best fit the changes of the surface abundances 
are those computed with the vertical shear diffusion coefficient of Maeder (1997) and 
the horizontal shear diffusion coefficient by Zahn (1992).}

%\texttt{Rotation of stars has many interesting and important consequences on 
%the photometric and chemical evolution of galaxies. Many predictions of rotating stellar 
%models are now compared with observations of surface abundances and velocities, 
%with interferometric studies of fast rotating stars, with
%internal rotation profiles as they can be deduced by asteroseismology, just to cite only a few observational constraints. 
%In this paper, we investigate how the outputs of models depend on the prescriptions used
%for the diffusion coefficients involved into the shellular rotating models. After a recall of the various 
%prescriptions found in the literature, we discuss their impacts on the evolutionary tracks and lifetimes during the Main-Sequence (MS) phase, 
%the changes of the surface composition and velocities during the MS phase,
%the distribution of the core helium lifetime in the blue and the red part of the HR diagram,
%the extensions of the blue loops,
%the evolution of the angular momentum of the core during the core H and He-burning phases,
%and the synthesis of primary nitrogen in fast rotating metal poor massive stars. 
%While some of these outputs are little dependent on the prescriptions used
%(as for instance the evolution of the surface velocities), most of them present a significant dependance. 
%the models which the best fit the changes of the surface abundances are those computed
%with the vertical shear diffusion coefficient of Maeder (1997) and the horizontal shear diffusion coefficient by
%Zahn (1992).} 

\section{Rotation in stellar models}

In recent years, many effects of axial rotation on the structure and the evolution 
of massive stars have been studied \citep[see e.g. the recent review by][]{RMP2012}. 
Among the effects which are the most important are those linked to the transport 
of angular momentum and of chemical 
species in the interior of stars.
These may strongly affect 
many outputs of stellar models such as the 
variation with the age of the surface abundances and velocities, the evolutionary 
tracks and lifetimes, the nature of the supernova events and of the stellar remnants, 
and the nature and the amount of new synthesized species.
As a consequence, when results of rotating 
models are used in population synthesis 
models or in models for the chemical evolution of galaxies, quite different 
results are obtained with respect to results obtained from non-rotating 
models.

Most, if not all of the recent grids of rotating models have been computed 
in the framework of the theory proposed by \citet{Zahn1992}, 
with further improvements by \citet{Maeder1998}. This was 
named the theory of 
shellular rotation, since it is based on the assumption that on an isobaric surface, 
the angular velocity, $\Omega$, is nearly constant, that means that any variations can 
be considered as a small perturbation. This nearly constant value of $\Omega$ on an 
isobaric surface is due to the fact that along those directions, there are neither
stable temperature nor density gradients which 
counteract shear turbulence. This implies the existence of strong ``horizontal'' (\ie along an isobaric surface) 
diffusion coefficient called $D_{\rm h}$ hereafter. In the following, when we speak about 
rotating models, we implicitly assume that 
we consider models with shellular rotation.
The present models do not include the effects of the dynamo theory suggested by 
\citet{Spruit2002}\footnote{Numerical simulations by
\citet{Zahn2007} have studied MHD instabilities arising in the radiation zone of a differentially rotating star, in which a poloidal field of fossil origin is sheared into a toroidal field. Their simulations show no sign of dynamo action.}.

In the framework of the shellular theory of rotation, the 
equation describing the transport of chemical species is a 
pure diffusive equation \citep{Chaboyer1992} written
\begin{eqnarray}
  \varrho{\partial X_i \over \partial t}\bigg|_{M_r} = {1 \over r^2} \, {\partial \over \partial r}
  \left (\varrho \, r^2 \, D_{\text{chem}} \, {\partial X_i \over \partial r}\right )  \; ,
\label{difx}
\end{eqnarray}
\noindent
where $X_i$ is the abundance in mass fraction of particles $i$, and $D_{\text{chem}}$, 
the appropriate diffusion coefficient for chemical elements (see
below).

In a differentially rotating star, the evolution of the angular velocity $\Omega$ 
has to be followed at each level $r$ (for shellular rotation), so that a full 
description of $\Omega(r,\, t)$ is available. The values of $\Omega(r,\, t)$ influence 
the mixing of elements and in turn the evolution of $\Omega(r,\, t)$  also depends 
on the mixing processes and on the distribution of 
the elements. The derivation of the equation for the transport of angular momentum is 
not straightforward. In the case of shellular rotation, the equation in the Lagrangian 
form becomes \citep{Zahn1992, Maeder2009}
\begin{eqnarray}
\varrho \, {\partial \over \partial t}(r^2 \bar{\Omega})_{M_r}
={1 \over 5 \, r^2}{\partial \over \partial r}(\varrho \,  r^4 \bar{\Omega} \,U_2(r))
+{1 \over r^2}{\partial \over \partial r}
\left(\varrho  \, D_{\text{ang}} \,  r^4 \,  {\partial \bar{\Omega} \over \partial r} \right) \; .
\label{eqn7}
\end{eqnarray}
\noindent
Here, $\bar{\Omega}$ is the average value of $\Omega$ on an 
isobar. $U_2$ is the radial component of the meridional circulation velocity, and 
$D_{\text{ang}}$, the appropriate diffusion coefficient for angular momentum.
The second term on the right is a diffusion term, similar in its form  to
(\ref{difx}), while the first term on the right is  \emph{an advective term}, \ie
modeling
the transport by a velocity current. We notice that Eq.~(\ref{difx}) does not 
contain such an advective term. It could contain a term of that kind, however 
it can be shown  \citep{Chaboyer1992} that the combined effect of turbulence and 
circulation currents is equivalent  to a diffusion \emph{for the element transport} 
(see Eq.~\ref{difx}).

In the equation for the transport of chemical species in radiative zones, the 
diffusion coefficient, $D_{\text{chem}}$, is made up of two terms. These are the
vertical shear diffusion coefficient $D_{\rm shear}$ and the effective diffusion 
coefficient, $D_{\rm eff}$, which account for the resultant 
effect of the strong horizontal shear diffusion, $D_{\rm h}$ (\ie the shear on an 
isobaric surface), and of the meridional currents. 

In the equation for the transport of angular momentum, the diffusion coefficient, 
$D_{\text{ang}}$, is made up of only one term, the shear diffusion coefficient 
$D_{\rm shear}$.

%The expression for $D_{\rm eff}$ is taken equal to 
%while the expression for $U_2(r)$ is given by expression (NN) in REF.

For the coefficients $D_{\rm shear}$, we can find two different expressions in the 
literature.
\begin{description}[font=\bfseries]
\item[$D_{\rm shear}$ from Maeder (1997, M97) \label{maed97}]
  \begin{equation}
    D_\text{shear} =  f_\text{energ} \frac{H_P}{g\delta}\frac{K}{\left[\frac{\varphi}{\delta}\nabla_\mu 
        + \left( \nabla_\text{ad} - \nabla_\text{rad} \right)\right]} 
    \left( \frac{9\pi}{32}\ \Omega\ \frac{\text{d} \ln \Omega}{\text{d} \ln r} \right)^2
  \end{equation}
  where $K = \frac{4ac}{3\kappa}\frac{T^4\nabla_\text{ad}}{ \rho P \delta}$, and with  
  $f_\text{energ} = 1$, and 
  $\varphi = \left( \frac{\text{d}\ln\rho}{\text{d}\ln\mu} \right)_{P,T} = 1$.
\item[$D_{\rm shear}$ from Talon \& Zahn (1997, TZ97)]
  \begin{equation}
    D_\text{shear} =  f_\text{energ} \frac{H_P}{g\delta}\frac{\left(K+D_\text{h}\right)}{\left[\frac{\varphi}{\delta}\nabla_\mu\left(1+\frac{K}{D_\text{h}}\right) + \left( \nabla_\text{ad} - \nabla_\text{rad} \right)\right]} \left( \frac{9\pi}{32}\ \Omega\ \frac{\text{d} \ln \Omega}{\text{d} \ln r} \right)^2
  \end{equation}
  with $K$, $f_\text{energ}$, and $\varphi$ as in (\ref{maed97}).
\end{description}
For the coefficients $D_{\rm h}$, we can find three different expressions in the 
literature:
\begin{description}[font=\bfseries]
\item[$D_{\rm h}$ from Zahn (1992, Z92)]
  \begin{equation}
    D_\text{h} =  \frac{1}{c_\text{h}}\ r\ \left| 2\,V(r) - \alpha\,U(r) \right| 
  \end{equation}
  where $\alpha = \frac{1}{2} \frac{\text{d} \ln (r^2 \bar{\Omega})}{\text{d} \ln r}$ and 
  $c_\text{h} = 1$.
\item[$D_{\rm h}$ from Maeder (2003, M03)]
  \begin{equation}
    D_\text{h} =  A\ r\ \left( r\bar{\Omega}(r)\ V\ \left| 2V-\alpha U \right| \right)^{1/3} \\
  \end{equation}
  with $\alpha$ as in Eq.~(5) and $A=0.002$.
\item[$D_{\rm h}$ from Mathis et al. (2004, MZ04)]
  \begin{equation}
    D_\text{h} =  \left( \frac{\beta}{10} \right)^{1/2} \left( r^2\bar{\Omega} \right)^{1/2} 
    \left( r \left| 2V-\alpha U \right| \right)^{1/2}
  \end{equation}
  with $\alpha$ as in Eq.~(5) and $\beta=1.5\cdot10^{-6}$.
\end{description}
All prescriptions use the same effective mixing coefficient for the chemical species:
\begin{equation}
  D_\text{eff} =  \frac{1}{30} \frac{\left| r\ U(r) \right|^2}{D_\text{h}}.
\end{equation}

There are therefore 6 different combinations of the two shear diffusion 
coefficients and of the three horizontal diffusion coefficients.
The physics sustaining the different expressions for these various diffusion 
coefficients is described in details in the papers indicated above and we shall 
not recall them in the present work. We just summarize below a 
few facts which are useful to keep in mind in order to understand their different 
impacts in stellar models.
\begin{itemize}
\item Since the angular momentum is transported mainly by the meridional currents, one can expect 
  that changing the expressions for the diffusion coefficients will have only a weak impact
  on the angular momentum distribution in stars. We shall see that this is well verified by 
  the numerical models.
\item The diffusion coefficient $D_{\text{eff}}$ is the key quantity determining the efficiency 
  of mixing in regions where there 
  is a strong $\mu$-gradient, for instance at the border of 
  the H-convective core.
\item The larger $D_\text{h}$ is, 
  the smaller will be $D_\text{eff}$, and thus less mixing 
  will occur in regions of strong $\mu$-gradients. The expressions of M03 and MZ04 for 
  $D_\text{h}$ are larger than the expression given by Z92.
\item The diffusion coefficient $D_\text{shear}$ is the key quantity determining the efficiency 
  of mixing in regions
  with weak or no $\mu$-gradients, typically in the radiative 
  envelope of massive stars above the H-convective core.
\item The two expressions of $D_\text{shear}$ are strictly equivalent in zones 
  with no $\mu$-gradients.
\item The ratio
  $D_\text{shear}\text{(M97)}/D_\text{shear}\text{(TZ97)} \sim K/D_\text{h}$ in regions where  
  ${\varphi \over \delta}\nabla_\mu$ is significantly larger than the difference 
  $\nabla_\text{ad} - \nabla_\text{rad}$. Since $D_\text{h}$ is inferior to $K$, one has that
  $D_\text{shear}\text{(M97)} > D_\text{shear}\text{(TZ97)}$.
\end{itemize}

\section{The models computed}

In the present work, we study the implications of these different choices on the following model outputs:
\begin{itemize}
\item Evolutionary tracks and lifetimes during the Main-Sequence (MS) phase.
\item Changes of the surface composition during the MS phase.
\item Changes of the surface velocities during the MS phase.
\item The distribution of the core helium lifetime in the blue and the red part of the HR diagram.
\item The extensions of the blue loops.
\item The evolution of the angular momentum of the core.
\item The synthesis of primary nitrogen in fast rotating metal poor massive stars.
\end{itemize}
For that purpose we have computed models for different initial masses, metallicities, 
rotations with each one of the 6 possible combinations of values for ($D_\text{shear}, D_\text{h}$) 
(see Table~1). 
For each mass and metallicity, the models are labeled by one digit and one letter: 1 is 
for models computed with the shear diffusion coefficient of Maeder (1997), and
2 for models computed with the shear diffusion coefficient of Talon \& Zahn (1997), 
the letters A, B and C are respectively for
the horizontal diffusion coefficient from Zahn (1992), Maeder (2003) and Mathis et al. (2004).

In Table 1, the first column gives the prescription used. The time-averaged equatorial 
velocity during the MS phase is given in column 2, the MS lifetime is given in column 3,
the difference between the surface helium abundance in mass fraction  at the end of the 
MS phase and on the ZAMS is given in column 4.
Column 5 presents the N/H ratio obtained at the surface, at the end of the MS phase, 
and normalized to the initial N/H value. The core He-burning lifetime, and the analogs
of columns 4 and 5 but at the end of the core He-burning phase are indicated in 
columns 6, 7 and 8 respectively. The duration of the core He-burning phase
spent in the red ($\log T_{\rm eff} < 3.68$), in the blue ($\log T_{\rm eff} > 3.87$) and 
in the yellow part ($3.68 < \log T_{\rm eff} < 3.87$) of the HR diagram are given in 
columns 9, 10 and 11 respectively. The ratio of the time spent
in the blue to that spent in the red is shown in column 12, the masses of the helium 
cores, of the carbon-oxygen cores and of the remnants are given in columns
13, 14 and 15. The mass of nitrogen produced divided by the mass of CNO elements 
initially present is given in column 16. 

\begin{table}
\caption{Models computed in the present work}
\label{tab:1}       % Give a unique label
%
% Follow this input for your own table layout
%
\scriptsize{
\scalebox{0.9}{%
\begin{tabular}{cccccccccccccccc}
\hline
      & $\overline \upsilon_{\rm eq}$ & $t_{\rm H}$ & $\Delta Y_s$ & ${(N/H) \over (N/H)_{\rm ini}}$ & $t_{\rm He}$ &   $\Delta Y_s$ & ${(N/H) \over (N/H)_{\rm ini}}$ & $t_{\rm blue}$ & $t_{\rm red}$ & $t_{\rm yel}$ & $t_{\rm blue} \over t_{\rm red}$ & M$_{\rm He}$ & M$_{\rm CO}$ & M$_{\rm rem}$ & $M(^{14}N) \over M(CNO_{\rm ini})$\\
      & km s$^{-1}$ & My & & & My & & & My& My & My & & M$_\odot$ & M$_\odot$ & M$_\odot$ & \\
\noalign{\smallskip}\svhline\noalign{\smallskip}      
\multicolumn{16}{c}{9 M$_\odot$, Z=0.002, $\upsilon_{\rm ini}/\upsilon_{\rm crit}=0.5$} \\    
\noalign{\smallskip}\svhline\noalign{\smallskip}
1A & 177         & 30.065        & +0.0149          &  4.9      & 2.966                 & 0.0157            & 4.9   &    3.0964   & 0.0970   & 0.0104   & 31.9   & 2.9 & 1.5 & 1.3 & 0.5     \\
1B & -0.6\%    & -11.8\%        & +0.0051          &  3.8      & +37.6\%             & 0.0055            & 3.8   &    +37.5\% & +21.0\% & +27.9\% & 36.3   & 2.7 & 1.5 & 1.3 & 0.5     \\
1C & +0.6\%    & -13.3\%        & +0.0024          &  3.1      & +28.9\%             & 0.0064            & 3.8   &    +28.6\% & +63.1\% & $\times$ 2 & 25.2   & 2.5 & 1.4 & 1.2 & 0.5     \\
2A & -1.1\%    & +0.3\%        & +0.0004          &  2.2      & -11.7\%             & 0.0123            & 3.5   &   -52.4\% & $\times$ 11.1 & $\times$ 27.3 & 1.4   & 2.9 & 1.5 & 1.3 & 0.4     \\
2B & -2.2\%    & -11.7\%        & +0.0003          &  3.0      & +17.4\%             & 0.0040            & 4.2   &   -10.4\% & $\times$ 9.6 & $\times$ 8.7 & 3   & 2.6 & 1.4 & 1.2 & 0.9     \\
2C & +0.6\%    & -13.3\%        & +0.0017          &  3.4      & +37.1\%             & 0.0099            & 4.8   &   +27.7\% & $\times$ 3.8 & $\times$ 3.3 & 10.8   & 2.5 & 1.6 & 1.3 & 0.4     \\
\noalign{\smallskip}\svhline\noalign{\smallskip}      
\multicolumn{16}{c}{9 M$_\odot$, Z=0.014, $\upsilon_{\rm ini}/\upsilon_{\rm crit}=0.5$} \\    
\noalign{\smallskip}\svhline\noalign{\smallskip}
1A & 163         & 31.423        & +0.0050          &  2.5      & 3.291                 & 0.0620            & 5.8   &    1.3982                    & 2.0974             & 0.0997                          & 0.7   & 2.5 & 1.4 & 1.2 & 0.4     \\
1B & -1.2\%    & -13.0\%        & +0.0009          &  1.8      & +47.6\%             & 0.0363            & 4.9   &    +64.0\%                  & +35.1\%           & -14.0\%                        & 0.8   & 1.6 & 1.4 & 1.2 & 0.3     \\
1C & -0.6\%    & -14.7\%        & +0.0003          &  1.6      & +36.1\%             & 0.0314            & 4.7   &    +53.8\%                  & +27.2\%           & -18.7\%                        & 0.8   & 1.4 & 1.2 & 1.1 & 0.4     \\
2A & -0.6\%    & -0.2\%        & +0.0001          &  1.3      & -5.3\%                  & 0.0495            & 4.5   &   -42.9\%                      & +12.5\%        & $\times$ 2.1                 & 0.3   & 2.5 & 1.4 & 1.2 & 0.3     \\
2B & -1.8\%    & -12.9\%        & +0.0000          &  1.6      & +17.5\%             & 0.0240            & 4.2   &   $\times {1 \over 9.2}$ & $\times$ 2.0 & 0.0                                & 0.04   & 1.7 & 1.1 & 1.1 & 0.4     \\
2C & -0.6\%    & -14.7\%        & +0.0002          &  1.7      & +36.8\%             & 0.0242            & 4.7   &   $\times {1 \over 8.8}$ & $\times$ 2.3 & $\times {1 \over 14.4}$ & 0.03   & 1.5 & 1.3 & 1.2 & 0.3     \\
\noalign{\smallskip}\svhline\noalign{\smallskip}      
\multicolumn{16}{c}{15 M$_\odot$, Z=0.002, $\upsilon_{\rm ini}/\upsilon_{\rm crit}=0.5$} \\    
\noalign{\smallskip}\svhline\noalign{\smallskip}
1A & 197         & 13.173        & +0.0248          &  5.4      & 1.323                 & 0.0612            & 6.9   &    1.2622                          & 0.0749             & 0.0416                          & 16.9     & 5.3 & 3.0 & 1.7 & 0.4     \\
1B & -3.0\%    & -2.2\%        & +0.0215          &  5.1      & +8.8\%                & 0.0237            & 7.1   &    +18.6\%                        &  0.0                   & $\times {1 \over 3.5}$   & -            & 4.8 & 2.8 & 1.6 & 0.5     \\
1C & -2.5\%    & -5.3\%        & +0.0105          &  4.2      & +12.9\%             & 0.0496            & 5.9   &    +24.6\%                         &  0.0                   & 0.0                             & -             & 4.5 & 2.6 & 1.6 & 0.5     \\
2A & -0.5\%    & -0.4\%        & +0.0012          &  2.4      & -17.5\%              & 0.0549            & 5.8   &   $\times {1 \over 34.2}$   &  $\times$ 4.5     & $\times {1 \over 2.7}$ & 0.03   & 5.8 & 3.6 & 1.8 & 0.4     \\
2B & -3.0\%    & -11.5\%        & +0.0006          &  3.1      & +1.8\%             & 0.0110            & 4.3   &   +6.3\%                           & -35.1\%              &  $\times {1 \over 2.0}$  & 27.6      & 4.9 & 2.8 & 1.6 & 0.4     \\
2C & -1.5\%    & -13.5\%        & +0.0039          &  3.9      & +43.9\%            & 0.0047            & 4.0   &  +53.9\%                         & $\times {1 \over 8.6}$ & $\times {1 \over 3.6}$ & 167.5   & 4.6 & 3.4 & 1.8 & 0.4     \\
\noalign{\smallskip}\svhline\noalign{\smallskip}      
\multicolumn{16}{c}{40 M$_\odot$, Z=0.00001, $\upsilon_{\rm ini}/\upsilon_{\rm crit}=0.75$} \\    
\noalign{\smallskip}\svhline\noalign{\smallskip}
1A & 702         & 5.559        & +0.0949          &  47.9      & 0.409           & 0.0964            & 48.2   &   0.4161    & 0.0      & 0.0       & -     & 17.1 & 14.0 & 4.3 & 1.1     \\
1B & -2.4\%    & -2.6\%        & +0.0743          & 44.2       & +5.7\%        & 0.0753            & 44.4   &    +6.0\%   &  0.0      & 0.0       & -     & 16.7 & 12.8 & 4.1 & 15.3     \\
2A & -2.8\%    & +1.4\%        & +0.0459          & 30.7      & -2.5\%         & 0.2006            & 50.9   &   -20.8\%   &  0.0     & 0.0748 & -      & 23.9 & 17.3 & 5.3 &  8.2     \\
2B & -5.3\%    & -7.4\%        & +0.0162          &  27.1      & $> 5.2\%$   & 0.0164            & 27.2   &   +4.8\%    &  0.0      & 0.0      & -      & 20.9: & 12.9: & 4.1: & 105.8:     \\
\noalign{\smallskip}\hline\noalign{\smallskip}
\end{tabular}
}}
\end{table}

\section{Evolutionary tracks and lifetimes during the Main-Sequence phase}

\begin{figure}
\sidecaption
% For example, with the graphicx style use
\includegraphics[scale=.43, angle=-90]{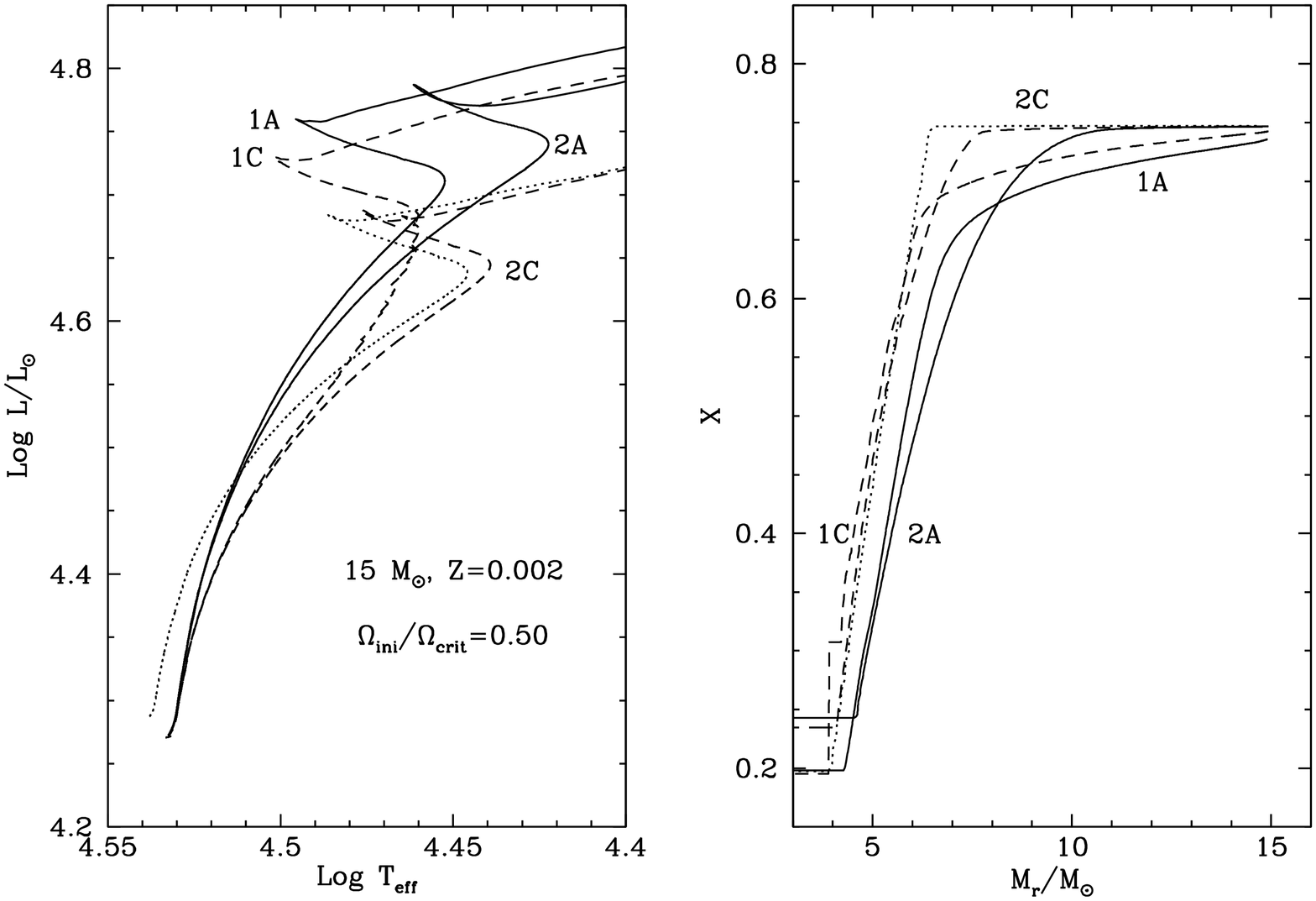}
%
% If no graphics program available, insert a blank space i.e. use
%\picplace{5cm}{2cm} % Give the correct figure height and width in cm
%
\caption{Evolutionary tracks in the Hertzsprung-Russell diagram (left panel) and the 
variation of the mass fraction of hydrogen as a function of the Lagrangian mass 
coordinate (right panel) for the 15 M$_\odot$ at $Z$=0.002 and with 
$\Omega_{\rm ini}/\Omega_{\rm crit}=0.5$. The dotted line corresponds to the 
non-rotating model, the other models are labeled as indicated in Table~1. The models 
used in the right panel have all a $\log T_{\rm eff}$ equal to 4.48. The central 
hydrogen mass fraction is equal to 0.1973 (non rotating model), 0.1982 (1A), 0.2068 (1C), 
0.2427 (2A), 0.2244 (2C). The models 1B and 2B  are not shown since they present many 
similarities with the models 1C and 2C.} 
\label{fig1}       % Give a unique label
\end{figure}

Figure~1 presents the tracks in the Hertzsprung-Russell diagram 
(left panel) and the variation of the mass fraction of hydrogen as a function of the 
Lagrangian mass 
coordinate (right panel) for the 15 M$_\odot$ at $Z$=0.002 and with 
$\Omega_{\rm ini}/\Omega_{\rm crit}=0.5$. One can classify the models in three categories:
\begin{enumerate}
\item The models which present very little differences with respect to the non-rotating 
  model. These are the models 2B (not shown in Fig.~1)  and 2C.
\item The models which become overluminous with respect to the non-rotating model but 
  present no significant extension towards lower effective temperatures. These are the
  models 1A, 1B (not shown) and 1C.
\item Only one model, the model 2A, becomes overluminous and reaches lower effective 
  temperatures at the end of the MS phase.
\end{enumerate}

This behavior reflects differences in the efficiency of mixing in different regions 
of the stars. 
The models which are of the less efficiently mixed 
stars
(2B and 2C, due to large $D_\text{h}$ 
and thus small $D_\text{eff}$) show indeed very little difference
in their non-rotating tracks (compare lines for the model 2C and the non-rotating model 
in Fig.~\ref{fig1}).
The model 2A presents a situation where there is an efficient mixing at the border of 
the convective core but where the shear is not so efficient in the radiative envelope. 
Thus the surface abundances are not yet modified at the stage represented in 
Fig.~\ref{fig1} (right panel). The larger core increases the luminosity and also is 
responsible for the extension towards lower temperature of the MS band in the HR diagram 
(an effect similar to an extension of the core produced by an overshoot).
Finally the model which presents the greatest efficiency of mixing both at the border 
of the core and in the radiative zone is the model 1A. This  mixing
keeps this model in bluer positions in the HR diagram compared to 
model 2.

One notes that in model 1A, the $\mu$-gradient (with respect to mass) is steeper 
than in the model 2A, while both models have the same 
expression for $D_{\rm eff}$ (since they have the same expression for $D_\text{h}$). 
This is because in model 1A, which 
has $D_{\rm shear}$ larger than in 2A, 
hydrogen flows more efficiently inwards and helium outwards. This replenishes
hydrogen at the border of the core. The net effect of these diffusions of 
hydrogen and helium is to make the star more luminous and  bluer. The same occurs 
in the 1C model, although the effect is less marked  
because of the smaller value for $D_{\rm eff}$.

In Table~1, the MS lifetimes are indicated in column 3. 
The first row for each model gives the value of the MS 
lifetime in million years for the model 1A. The rows for the other models
show the differences in percentage with respect to the 
value obtained for the model 1A.
We see that the impact on the MS lifetimes remains modest (at most 13.5\% for the 
15 M$_\odot$ models considered here) 
with respect to the precision with which an age estimated can be made
through the fitting of an isochrone in this mass domain. On the other hand the 
scatter is not negligible with respect to the amplitude of the effect
of the increase of the MS lifetime due to rotation. Indeed, the 
increase of the MS lifetime with respect to the non-rotating case amounts 
to 18.5\% for the 15 M$_\odot$ model.

All the models computed here
account for the same overshoot, but Fig.~\ref{fig1} shows that the track will 
present quite different extensions due to the various prescriptions. One sees also
that the models 1A and 2A (those computed with the smaller $D_\text{h}$ and thus 
greater $D_{\rm eff}$) are the only ones presenting
an extension of the core with respect to the non-rotating 
model. Therefore only the use 
of these two prescriptions can attribute part of the extension of the convective 
core to an effect of rotational mixing. Let us note that the recent determination 
of the extension of the mixed core in fast rotating stars by \citet{Neiner2012} 
seem to support the view that rotation enlarges the convective core. This would 
support prescriptions 1A or 2A.
%Therefore, as already emphasized by \citet{TZMM1997}, one sees 
%that the extension of the MS band is not completely due to
%the extension of the convective core only but also on the degree of mixing in the radiative envelope. This is of course quite natural, it suffices
%to think about the case of homogeneous evolution to realize that. The models presented here with some mixing in the radiative envelope are
%intermediate cases between the classic evolution (no mixing in the radiative envelope) and complete mixing of the whole star.
In that case, one should use slowly rotating stars in order to constrain the extension of the core
due to the process of convective penetration alone as is done, for instance, in \citet{Ekstrom2012}. 

\section{Changes of the surface composition during the MS phase}

\begin{figure}
\sidecaption
% For example, with the graphicx style use
\includegraphics[scale=.40, angle=0]{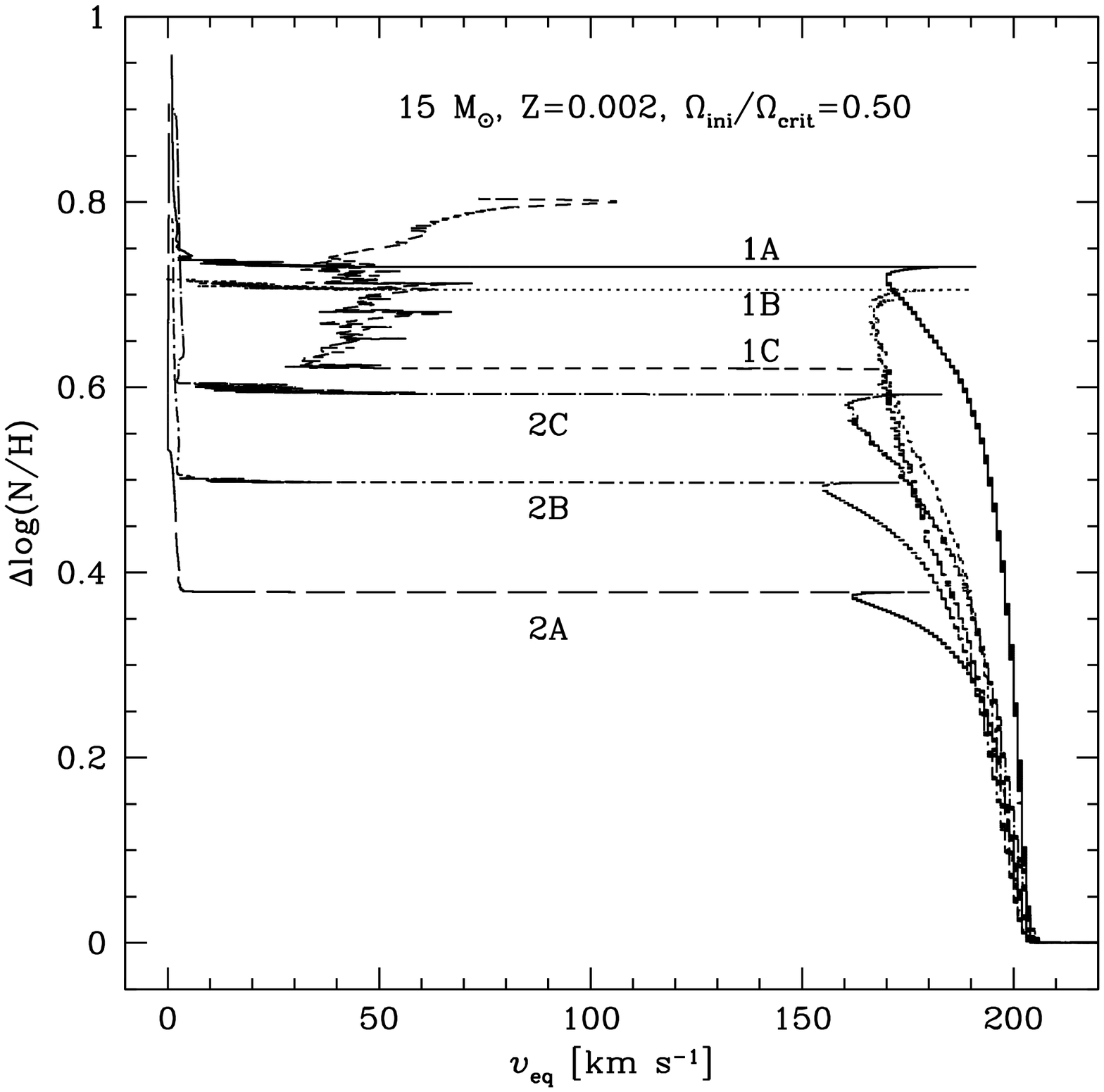}
\hfill
\includegraphics[scale=.40, angle=0]{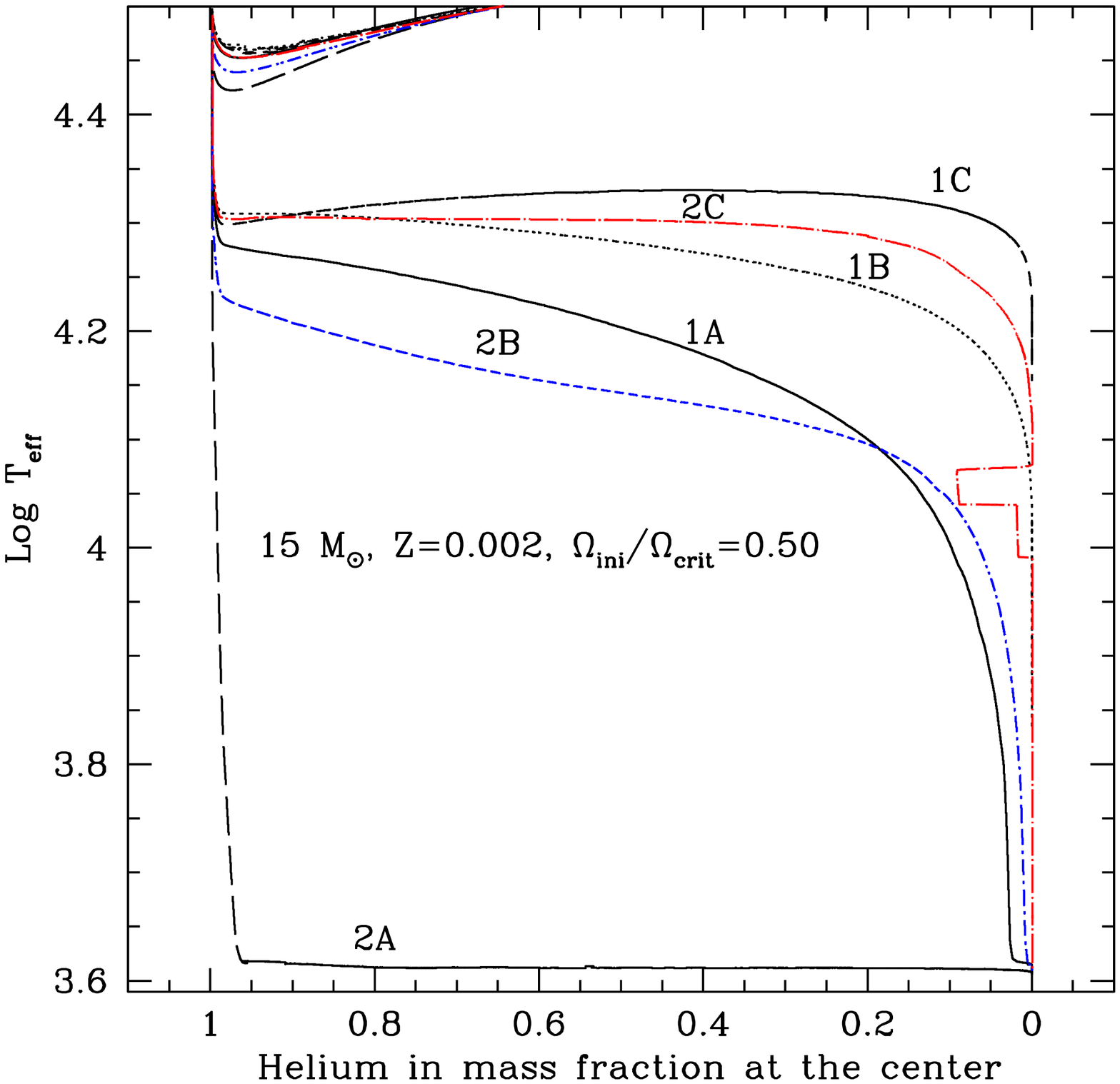}
%
% If no graphics program available, insert a blank space i.e. use
%\picplace{5cm}{2cm} % Give the correct figure height and width in cm
%
\caption{{\it Upper panel} Evolution of the N/H ratio at the surface of stellar models 
computed with various diffusive coefficients as a function 
of the equatorial velocity. {\it Lower panel} Evolution of the effective temperature 
during the core He-burning phases for different prescriptions
of the diffusive coefficients.} 
\label{fig2}       % Give a unique label
\end{figure}

The changes of the surface composition during the MS phases can be seen in Table~1 
and in Fig.~\ref{fig2} for the 15 M$_\odot$ at $Z$=0.002 and $\Omega_{\rm ini}/\Omega_{\rm crit}=0.5$
cases.
Column 3 and 4 of Table~1 give, at the end of the MS phase, the excesses of helium at 
the surface of the star (in mass fraction and with respect to the initial value) 
and the ratio of nitrogen to hydrogen normalized to the initial value, respectively.

Independent of the prescriptions used, one notes, as was already obtained in previous 
works \citep[see][]{Maeder2001} that the surface enrichment in nitrogen increases in 
increasing initial stellar masses and that it also 
increases for decreasing metallicity, in both 
cases keeping the initial rotation the same and comparing stars at similar evolutionary 
stages.

One sees also that the surface enrichments are higher when $D_{\rm shear}$ is higher 
(compare models of series 1 with models of series 2). This is quite logical since 
$D_{\rm shear}$ is the parameter which governs the transport of
chemical elements in the region extending from the vicinity of the core up to the 
surface.

The greatest surface enrichments are always obtained for the model 1A, the smallest 
for the model 2A (except for the very metal poor 40 M$_\odot$ model, but the 
difference between models 2A and 2B is quite small). 
For a given model, the variation of the prescriptions used produces a scatter of 
the N/H value obtained at the end of the MS phase of around a factor of 2. For the 
excesses in helium, the factors are greater, however, except for the 
40 M$_\odot$ at $Z$=0.00001 case, the enhancements are very modest and well 
below those which could be estimated from observed 
spectra.

\section{Changes of the surface velocity during the MS phase}

Column 2 in Table~1 compares the time-averaged equatorial velocity during the MS phase. 
The first row of the table indicates the equatorial surface velocity for model 
1A in km\,s$^{-1}$. The other lines indicate the difference in percentage
with respect to model 1A.
One sees that the differences are always less than 5.3\%, which is small. This is due to the 
fact that changing the prescriptions for the diffusion coefficients has only a very weak 
impact on the velocity of the meridional currents. Thus the transport of the angular moment
which is mainly driven by these currents is almost unaffected 
by changes of the diffusion coefficients.

This can also be seen in Fig.~2 (left panel), which shows the variation of the nitrogen 
enhancement with respect to the surface equatorial velocity (tracks on this diagram go 
from right to left as time proceeds). One sees that during the MS phase, 
all of the models span the same interval in equatorial velocities. What changes
is the surface nitrogen enrichment obtained at the end of the MS phase. 
In the plane N/H versus $\upsilon_{\rm eq}$, model 1A will 
produce a steeper relation
than model 2A. 

If one considers mean values of the ratio N/H for B dwarfs in the Galaxy and 
in the SMC one obtains enhancement factors between 1.6 and 2.5 for the Galaxy and 3.2 
and 6.3 for the SMC \citep[see Table~2 in][and references therein]{RMP2012}. These mean 
values should be obtained for stars with an average rotation during the MS phase of 
around 200 km\,s$^{-1}$. 

At solar metallicity, the 15 M$_\odot$ model 1A well reproduces the observed 
enrichments \citep{Ekstrom2012}. This is not a surprise, since the value of the 
parameter $f_{\rm energ}$ in $D_{\rm shear}$, chosen equal to one, has been
selected in order to fit these observed enrichments. With the same value of 
$f_{\rm energ}$, the 15 M$_\odot$ at Z=0.002 with an average rotation of  
200 km s$^{-1}$ and prescriptions 1A predicts an enhancement factor between 
3 and 5.3 in the last third of its MS lifetime, which is in the range
of the observed values. In that case, no calibration has been made and the 
good fit supports this kind of model. The other 
prescriptions gives too low surface enrichments keeping $f_{\rm energ}$ equal 
to 1.

We note that a value equal to 1 for $f_{\rm energ}$ implies that 
we really account for the physics involved into the expression  
for $D_{\rm shear}$, which would not be the case if one would have to multiply
the expression by a constant much greater or much smaller than one! 

\section{The distribution of the core helium lifetime in the blue and the red 
part of the HR diagram}

The observed number of blue to red supergiants in clusters at different metallicities 
is an important feature that stellar models should be able to reproduce. 
This is important for many reasons, for instance to predict the 
correct photometric evolution of young starburst regions or the nature of the 
progenitors of the core collapse supernovae. It happens that
observation shows that the blue to red ratio in clusters with masses at the 
turn off  between 9 and 30 M$_\odot$ increases when the metallicity increases, while 
standard stellar models predict that the blue to red supergiant ratio decreases 
when the metallicity increases \citep{Meylan1982, Langer1995, Eggenberger2002, BSRSL2011}.
The blue to red supergiant ratio
has also been 
discussed in the 
context of field stellar populations in
\citet{Hartwick1970, SKILL2002}. 

At the moment, there is no explanation for this general trend. On the other hands, 
many works could reproduce the blue to red supergiant ratios observed at one given 
metallicity by changing the
mass loss rates \citep[see e.g.][]{Salasnich1999} or mixing \citep{Langer1995, Maeder2001}. 
In this paper we shall not
discuss all the aspects of this question but focus on the importance of mixing. 

Looking  at the right panel of Fig.~\ref{fig2}, which shows the evolution of the 
effective temperature during
the core Helium burning phase, we see that only one set of diffusion coefficients (the one using
the D$_{\rm shear}$ of Talon \& Zahn 1997, and the D$_{\rm h}$ of Zahn 1992) makes the 
15 M$_\odot$ at Z=0.002 evolve rapidly to the red part of the 
HR diagram after the MS phase.
This is the prescription that we used in \citet{Maeder2001},
where we suggested that rotational mixing could help a lot in reproducing the 
observed blue to red supergiant ratio in the Small Magellanic Cloud. In view of 
the present results we see that, while this conclusion might always be correct,
it is however quite dependent on the prescriptions used for the diffusion coefficients.

It is interesting to identify from the numerical experiments performed in this work, the conditions
which favor a rapid redward evolution at low metallicity. It does appear that two conditions
have to be satisfied: 1) The mixing at the border of the convective cores (both during
the H- and He-burning phases) have to be sufficiently efficient. Indeed, we see that any 
very strong values for D$_{\rm h}$, which prevent any strong mixing in regions with strong 
molecular weight gradient ($\mu$-gradient), also prevents the star 
evolving to the red phase. 
2) The mixing in the zones where the $\mu$-gradients are weak, namely the outer
part of the radiative envelope, should not be too strong, because any strong mixing 
there would make the star be more ``homogeneous'' and thus 
maintain a bluer position in the HR diagram.
As very often noticed in
the literature, we see that this red to 
blue evolution is a
feature which is very sensitive to
many physical ingredients of the models.
The fact that it depends on subtle changes of the efficiency of mixing in different 
regions of the star is just one illustration of this. 

Taken at face value, the set of diffusion coefficients D$_{\rm shear}$ from 
Talon \& Zahn (1997), and the D$_{\rm h}$ of Zahn (1992) appear as the most favored to 
explain the blue to red supergiant ratio at low metallicity.
However, other parameters ---
for instance the changes of the surface abundances expected for an averaged rotational velocity ---
are better fitted with the prescription 1A (keeping  $f_{\rm energ}=1$). 
Moreover, since mass loss, both during the MS phase and at the red supergiant phase, 
plays a key role in shaping the blue to red supergiant ratio 
\citep[see e.g. the discussion in][]{BSRSL2011}, it may be premature 
to use the observed variation of the blue to red supergiant ratio to constrain the 
prescription to be used.
Probably in order to make progress in this area of research two 
important points 
have first to be settled:
1) to distinguish, using observations of the 
surface abundances and/or of the vibrational properties,  those blue 
supergiants which are direct successors of MS stars from those which are on a blue loop after
a red supergiant stage. The ability to distinguish, 
at different metallicities,
between those blue supergiants coming from the MS phase from those coming 
from the red supergiant phase would improve 
considerably
our understanding of the blue supergiant formation process; 
2) to obtain more reliable mass loss rates especially during the
red and blue supergiant phases.

\section{The extensions of the blue loops}

\begin{figure}
\sidecaption
% For example, with the graphicx style use
\includegraphics[scale=.40, angle=0]{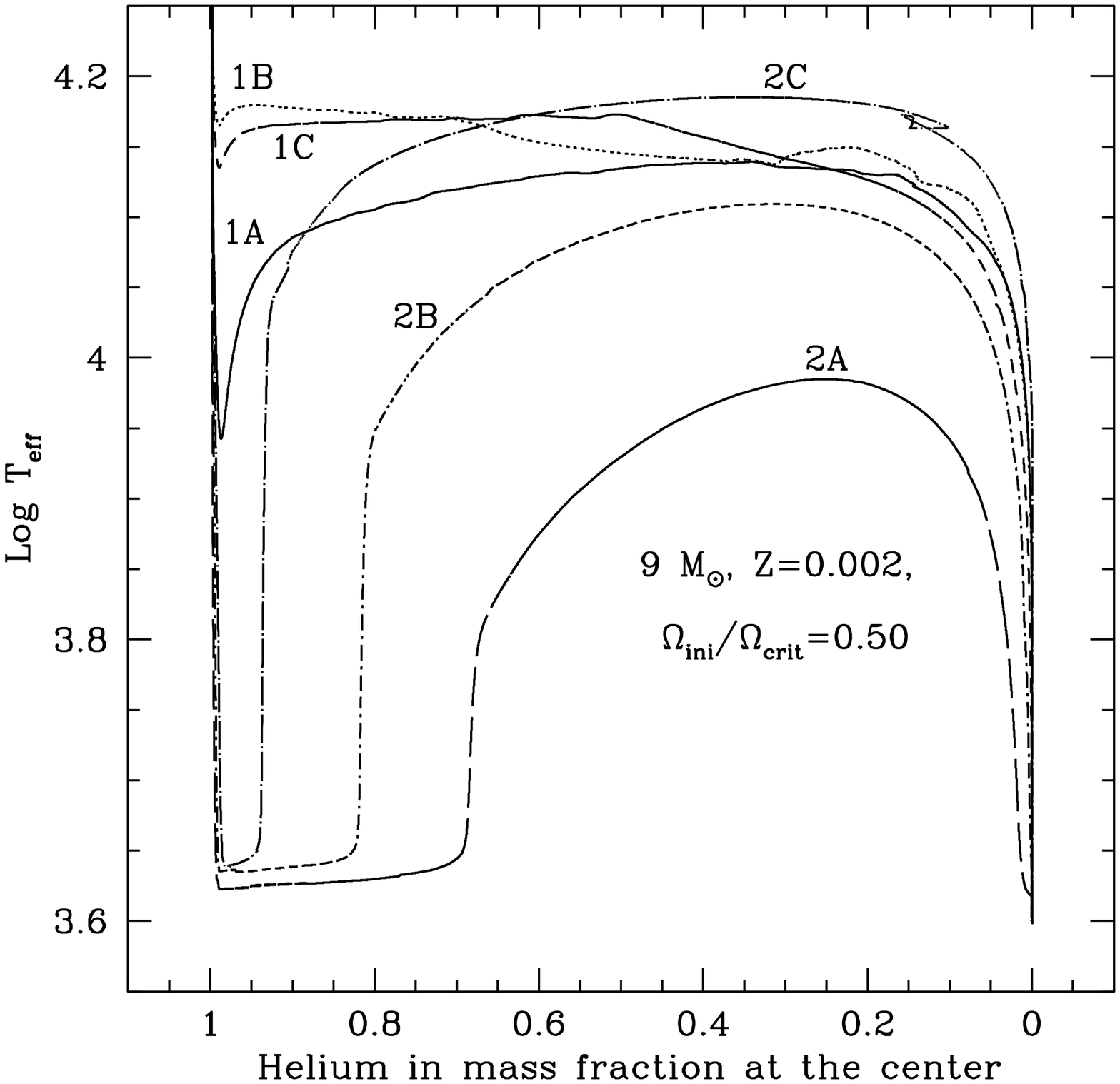}
\hfill
\includegraphics[scale=.40, angle=0]{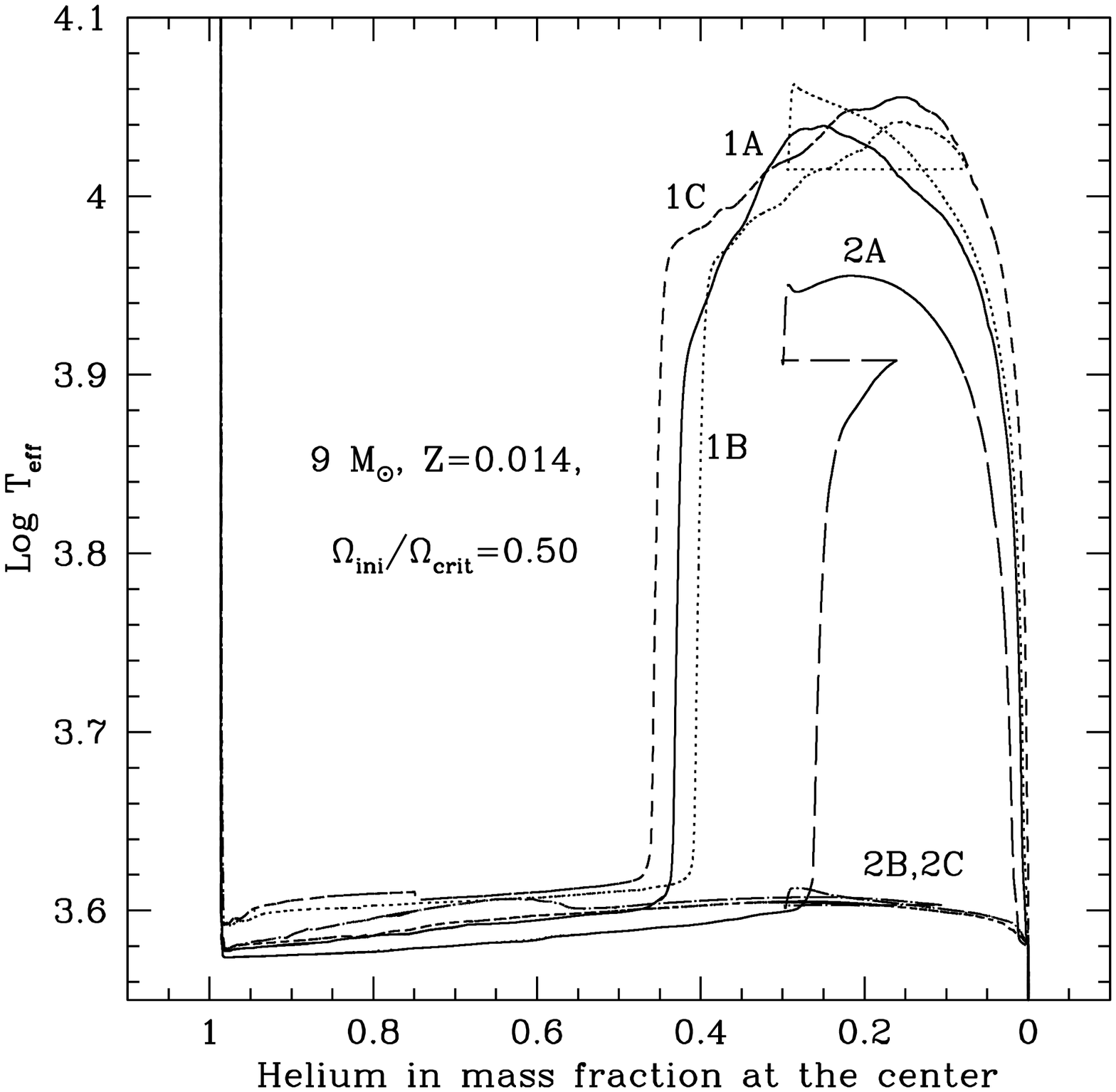}
%
% If no graphics program available, insert a blank space i.e. use
%\picplace{5cm}{2cm} % Give the correct figure height and width in cm
%
\caption{{\it Upper panel} Evolution of the effective temperature during the 
core He-burning phases for different prescriptions
of the diffusive coefficients in a rotating 9 M$_\odot$ stellar model at 
$Z$=0.002. {\it Lower panel} Same as the panel on the left at $Z$=0.014.} 
\label{fig3}       % Give a unique label
\end{figure}

Another feature which is sensitive to the form of the diffusion coefficients
is the extension of the blue loops for stars with masses between about
3 to 12 M$_\odot$. This is important in order to predict the populations of Cepheids, 
and also for
for the blue to red supergiant ratio discussed above, since the presence of a blue 
loop (when compared with the same model without blue loop)
will reduce the lifetime of red supergiant but increase that of the blue 
supergiant.

Looking at Fig.~\ref{fig3}, we can see the following features: 
at $Z=0.002$, the duration of the blue loops increases more and more passing 
from models 2A, to 2B, and then to 2C. When the D$_{\rm shear}$ is changed (model 1),
the ``loop'' (if we can still speak of a loop in this case) even begins in the 
blue part of the HR diagram.

At Z=0.014, the situation is quite different, first the loops in all models are 
significantly reduced, which is a well-known effect 
when the metallicity increases. Second, models 2B and 2C do not show any loops. 
In case such prescriptions would be adopted, then only slow rotators can show a 
blue loop and thus explain the existence of Cepheids.If such prescriptions were to be adopted, 
then only slow rotators could show blue loops and thus explain the existence 
of Cepheids.

We can note also that in the first set of models (labels beginning with one, 
\ie D$_{\rm shear}$ from Maeder 1997), the impact of changing D$_{\rm h}$ on blue 
loops is quite modest in both metallicities. In the second set of models, 
changing D$_{\rm h}$ has a strong effect. 

At low metallicity, models 2A for 9 M$_\odot$ would be the more helpful to 
reconcile the theoretical predictions with the observations as was
suggested from previous section which focused on 15 M$_\odot$ stellar models.
At solar metallicity, whatever model is considered, the blue to red supergiant 
ratios do appear too low (at most 1 while it is observed at around 3). 
At this metallicity, the problem may be at least partially cured by enhancing 
the mass loss rate during the red supergiant phase.

\section{The angular momentum of the core}

During the evolution of a star, the core loses some angular momentum, 
mainly due to the effect of meridional currents.
It happens that, in shellular rotating models
without interior magnetic field, 
these losses are not sufficient to explain the relatively 
long observed rotation periods of young pulsars \citep{Heger2004}. 
This is the reason why some authors have considered much stronger coupling between 
the core and the envelope by introducing a magnetic field which forces solid body 
rotation or near solid body rotation during the MS phase \citep{Heger2005}.
Here, we have not accounted for such a strong coupling.
A question however that we may ask is 
the extent to which this loss of angular momentum depends on the prescription 
used.
Are there any of these prescriptions which would significantly change 
the angular momentum contained in the core at the end of its evolution?

A priori, one would expect that the loss of angular momentum by the core due 
to the transport processes should be only slightly 
dependent on the various prescriptions because, as already 
stressed, 
during most of the stellar lifetime, angular momentum is transported
by the meridional currents whose velocities are only weakly dependent on 
the choice of D$_{\rm h}$. 
Let us, however, check this point in the numerical models that we have. 
Since we have not pursued the computation beyond the end of the core He-burning phase, 
we compare here the angular momentum of the core obtained at the end of the core helium 
burning phase. 
%The variations
%of the angular momentum contained inside a mass M$_r$ at the end of the core helium burning phase for the 
%15 M$_\odot$ at Z=0.002 are shown in Fig.~\ref{fig4} below. 
The masses of the remnant of the different 15 M$_\odot$ Z=0.002 models are between
1.6 and 1.8 M$_\odot$. The angular momentum which would be locked
into the 1.6 M$_\odot$ remnant supposing that no change does occur in the advanced 
phases of the evolution, would be between 0.93 and 1.14 10$^{50}$
g cm$^2$ s$^{-1}$ depending on the prescription used. The analog values in the case that
the remnant is 1.8  M$_\odot$ would be between 1.19 and 1.44  10$^{50}$ g cm$^2$ s$^{-1}$. 
So we see that these quantities present a scatter around their mean values of at 
most 20\%.

Is such a scatter important? As it concerns the missing angular 
momentum loss of the core, the answer is clearly no. To illustrate this, 
let us derive the following numerical estimate: if we lock an angular 
momentum content of 1 10$^{50}$ g cm$^2$ s$^{-1}$ in a neutron star, it would show a 
rotation period of about 0.1 ms, smaller by a factor between 4 and 7 than the critical 
periods for neutron stars which are between 0.44 and 0.65 ms as given by \citet{IC2012}.
The period is also smaller by two to three orders of magnitude than the observed periods of young pulsars 
which are between 20 and 100 ms 
\citep{Muslimov1996, Marshall1998}.
We can at least conclude
that the missing transport mechanism cannot be due to a 
particular choice of the diffusion coefficients for D$_{\rm shear}$ and D$_{\rm h}$, since 
whatever choice is made, the angular momentum content of the core is more or less the same
at the end of the core He-burning phase.
 
The angular momentum losses of the core may be underestimated either during the H and 
He-burning phases of the star and/or in the advanced phases and/or at the time of the 
supernova explosion and/or during the early years of the evolution of the new born 
neutron star. It may be that magnetic braking may play a role in this
context \citep{Meynet2011}.

\section{The synthesis of primary nitrogen in fast-rotating metal-poor massive stars}

\begin{figure}
\sidecaption
% For example, with the graphicx style use
\includegraphics[scale=.54, angle=0]{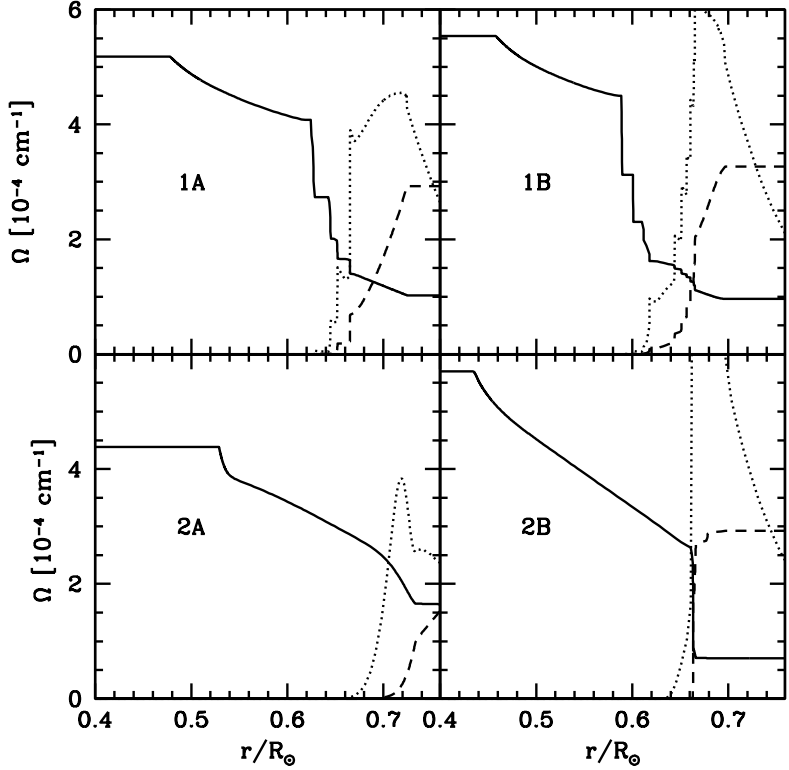}
%
% If no graphics program available, insert a blank space i.e. use
%\picplace{5cm}{2cm} % Give the correct figure height and width in cm
%
\caption{Variation of the angular velocity as a function of the radius 
inside the 40 M$_\odot$ stellar models when the mass fraction
of helium at the center is equal to 0.45. The dotted lines show, in each 
model, the variation as a function of the radius of the rate of production
of energy by H-burning, $\epsilon_{\rm H}$, in units of 
$10^5$ ergs g$^{-1}$ s$^{-1}$.} 
\label{fig4}       % Give a unique label
\end{figure}

Fast-rotating massive stars may be the sources of primary 
nitrogen in the early phases of the evolution of galaxies
\citep{Chiappini2006, Chiappini2008}.
Therefore it is important to assess the extent to which
the primary nitrogen production depends on the prescriptions used.
In the last column of Table~1, we have indicated the mass of nitrogen (in solar masses) 
present in the region outside the stellar remnant normalized by the mass of CNO elements 
that were initially present in the same region of the star. 
We call this quantity M($^{14}$N)/M(CNO).
When nitrogen is produced by the transformation of the carbon and oxygen initially present 
in the star (secondary nitrogen production channel), then the quantity
shown in Table~1 can at most be equal to one. 
It would be one if all the carbon and oxygen initially present in that 
region were to have been transformed into nitrogen. Actually it is 
less than one because, 
in the regions inside the H-burning shell, the nitrogen which has been
produced by the CNO burning will then have been further 
transformed into $^{22}$Ne.

We see that M($^{14}$N)/M(CNO) is inferior to one except 
in the fast-rotating very metal-poor 40 M$_\odot$ stellar models.
This illustrates the results already discussed by  \citet{paperVIII2002, Meynet2006} 
that rotational mixing, by bringing carbon and oxygen 
(freshly synthesized in the helium-burning core) into the H-burning shell, 
enhances 
significantly the quantities of nitrogen produced. The quantities produced are 
no longer limited by the initial metallicity of the star, since the carbon and oxygen 
transformed into nitrogen are synthesized by the star itself through helium transformation 
(primary nitrogen production channel). 

We see, however, that the enhancements of primary nitrogen production present great 
variations depending on the prescriptions used for D$_{\rm shear}$ and D$_{\rm h}$. 
Before we analyze these results in more 
detail, let us recall a few general facts: 
\begin{enumerate}
\item Primary nitrogen production depends on the efficiency of transport mechanisms 
   in the region between the He-burning core and the H-burning shell;
\item The diffusion coefficient which dominates the transport in that region is 
   D$_{\rm shear}$. 
A careful reader may be puzzled by such a statement since we mentioned above that 
D$_{\rm shear}$ operates mainly in weak $\mu$-gradient regions while the appropriate
diffusion coefficient in strong $\mu$-gradient regions would be D$_{\rm eff}$. Does it 
mean that the $\mu$-gradients are not so strong at the border of the He-burning core? 
The answer is yes. Indeed, in the core-shell intermediate region, most of the time, 
the gradients of $\mu$ are not very strong since the connected regions are all 
helium rich, moreover the gradient of $\Omega$ which enters into the expression 
for D$_{\rm shear}$ is important. 
%during the core helium burning phase, the situation with respect to the mu-gradients (which tends to becomes smaller due
%to less steep gradients in helium) and to the angular velocity gradients (which tends to become steeper)
%are such that the   D$_{\rm shear}$ dominates the mixing even at the border of the core. This is constrast with what happens during the core H-burning phase
%where the $\mu$-gradient at the border of the core makes the D$_{\rm shear}$ quite small and makes the D$_{\rm eff}$ the dominant
%factor in that region. 
\item Since D$_{\rm shear}$ is the dominant diffusion coefficient,
   the gradient of $\Omega$ becomes the key factor for primary nitrogen production, 
   a small gradient producing less efficient shear mixing than a steep gradient. 
\end{enumerate}

In Fig.~\ref{fig4}, we can see the variations of $\Omega$ as a function of the 
radius in the 40 M$_\odot$ stellar models when the mass fraction of helium
in the core is equal to 0.45 (that means models at the middle of the core 
He-burning phase). We see that models using the prescription by Maeder (1997)
have a smooth gradient immediately above the convective core and show in 
outer zones are a succession of regions with very steep and flat gradients.
We shall call this zone the cliff. The cliff corresponds to the transition between 
the envelope and the core. The H-burning shell is more or less at its base (see Fig. 4).

The smooth gradient results, at least in part, from the activity of the shear 
transport having occurred during the core He burning phase.
The flat $\Omega$ regions in the cliff are produced by intermediate convective 
zones which are no longer present at the stage shown in the figure but have 
appeared in previous stages of the evolution of the star. In the present models, 
we assume that convective regions rotate as solid bodies, hence the flattening of 
$\Omega$ in these zones. 
The flat regions present a very low D$_{\rm shear}$ coefficient since the 
$\Omega$-gradients are very small.
The part of the $\Omega$ profile which is important for primary nitrogen production is the 
portion with a smooth gradient
just above the He-burning core.

In the model 1B (higher D$_{\rm h}$ and thus smaller D$_{\rm eff}$), we note that the 
smooth gradient zone is more compact, making the gradient of $\Omega$ in that region 
steeper. This results from the less efficient chemical mixing 
at the border of the H-burning core. 
As a result, the helium core will be smaller and the transition zone between the core 
and the shell in a deeper part of the star.
Since model 1B presents a steeper gradient of $\Omega$ in the smooth gradient zone, 
it produces more primary nitrogen than model 1A (see Table~1).

%Such an $\Omega$ structure will  to primary nitrogen production
%for two reasons: first the smooth gradient produce a relatively slow mixing, second the region with flat gradients (which corresponds to the apparition in a relatively recent past of
%small intermediate convective zones)  stops the mixing in this region since there is no $\Omega$-gradient. This prevent the dredging up of carbon and oxygen in
%more active parts of the H-burning shell.

The configurations presented in the bottom panels, resulting from prescriptions 2A and 2B,
show different characteristics with respect to models 1A and 1B in upper panels: 
1) just above the core the gradient of $\Omega$ is steeper; 
2) the region with a succession of strong $\Omega$-gradients and flat portions 
    no longer exists.
These two features result from less efficient shear transport at the border of the burning cores.
Again, as in model 1, we see that when high values of D$_{\rm h}$ are used, stronger 
gradients are obtained.

%This is a more favorable situation for primary nitrogen production because steeper gradients
%produces more efficient mixing and the absence of ``flat $\Omega$ profiles'' do not stop mixing to reach outer layers.

We can see from Table~1 that, actually, the primary nitrogen production is larger in 
the models 1B, 2A and 2B than in model 1A as can be expected 
from the line of reasoning above. An interesting conclusion of this discussion is that primary nitrogen production depends mainly 
of the gradient of $\Omega$ just above the helium burning core. 
%The factors which make it stronger
%are not too efficient transport mechanisms in that region during the core He-burning phase and small helium burning cores
%(therefore not too efficient mixing also above the H-burning core).

%The presence of numerous relics of small convective zone producing the flat profiles zones in the upper panels
%of Fig.~\ref{fig4} are also a limiting factor for primary nitrogen production. One can wonder what makes the apparition of these small convective zones?
%Probably this is due to the more efficient mixing produced by the D$_{\rm shear}$ of Maeder (1997) which replenish more efficiently the H-burning shell making it stronger
%and thus more prone to become convective at least for some short periods, but this requires some more investigations to be confirmed.

%\begin{figure}
%\sidecaption
% For example, with the graphicx style use
%\includegraphics[scale=.54, angle=0]{N.eps}
%
% If no graphics program available, insert a blank space i.e. use
%\picplace{5cm}{2cm} % Give the correct figure height and width in cm
%
%\caption{Variation of the abundances of hydrogen, helium, carbon, nitrogen (boldface line) and oxygen as a function of radius
%inside the 40 M$_\odot$ stellar models when the mass fraction
%of helium at the centre is equal to 0.45.} 
%\label{fig5}       % Give a unique label
%\end{figure}

\section{Conclusions}

We have studied the impact of various prescriptions on important outputs of rotating 
massive star models. The main conclusions are the following:
\begin{itemize}
\item The outputs of stellar models which show marginal
  dependence on the prescriptions used for D$_{\rm shear}$ and D$_{\rm h}$ are 
  the MS lifetimes, 
  the evolution of the surface velocities, 
  and the evolution of the angular momentum of the core.
\item The outputs of stellar models which show significant dependence on the 
  prescriptions used for D$_{\rm shear}$ and D$_{\rm h}$ are 
  the shape of the evolutionary tracks, 
  the surface enrichments predicted for a given initial rotation, 
  the blue to red evolution, 
  and the extensions of blue loops and the amount of primary nitrogen produced.
\item The general trends of 
  the increase of the mixing efficiency with 
  the increase of the initial mass of the star, 
  of its initial rotation 
  and with the decrease of the initial metallicity remain the same whatever
  the prescriptions used for D$_{\rm shear}$ and D$_{\rm h}$.
\end{itemize}

The hope would be, of course, to identify the most realistic diffusion coefficients 
on the basis of physical considerations.
From an analytic point of view, the difficulty is mainly in how to treat turbulence. 
This is reflected by the fact that in each of these expressions, a free parameter 
is present ($f_{\rm energ}$ in D$_{\rm shear}$ for instance).
At the moment this free parameter is chosen so as to allow models to 
reproduce one well identified observed feature (for instance, the surface enrichments 
observed at the surface of MS B-type stars at solar metallicity).
Thus a good fit with the observed feature which has been used to calibrate the models is not
a test of the models. The tests of the models are comparisons with other observed features 
such as comparisons with surface enrichments for other velocities, 
metallicities or initial mass ranges. 
We show above that prescription 1A can account for the observed enrichments in the 
SMC, while the prescription has been calibrated on solar metallicity stars. Thus 
this can be viewed as a support of this kind of model.

We see also that one set of prescriptions 
cannot give satisfactory fit to all the observed features discussed here. 
This is expected since these observed features are not all governed only 
by the way mixing is treated. As indicated in the paper, the blue to red 
supergiant ratio, for instance, also depends a lot on the way in which
the mass loss due to stellar winds is implemented in the models. 
Moreov er, other effects, not accounted for here, such as close 
binary evolution, or magnetic braking, may also contribute to some of the 
observed features.

In order to progress on the theoretical side, the key points reside in the capacity 
to treat turbulence in a more rigorous way. The key to progress on the theoretical
front would seem to be the ability to treat turbulence in a more rigorous way.
Probably multidimensional hydrodynamical simulations can provide important hints 
on this topic \citep{Arnett2011}. 

Another complementary approach is to use well-observed stars to constrain the models. 
Stellar models which can account for many observed features have a great chance 
to provide a realistic description of the structure of stars. It may be sometimes for 
wrong reasons, in the sense that the physical process 
invoked may not be  exactly the one operating in nature, but  the effect of this 
physical process, whatever it is, has some chance to produce the structure as obtained 
in the model which best fits the observations.

We are confident that improvements in the two directions of hydrodynamical simulations 
and comparisons of evolutionary models with observed stars will allow us
to constrain the possibilities for the physics occurring 
in stellar interiors. 

\acknowledgement{The authors thank Dr K\'evin Belkacem for the careful copy editing of the
manuscript.}

\bibliographystyle{aa}
\bibliography{MyBiblio}

\end{document}